\documentclass [sigconf]{acmart}
\AtBeginDocument{%
  }

\copyrightyear{2025}
\acmYear{2025}
\setcopyright{rightsretained}
\acmConference[CHItaly 2025]{16th Biannual Conference of the Italian SIGCHI Chapter}{October 06--10, 2025}{Salerno, Italy}
\acmBooktitle{16th Biannual Conference of the Italian SIGCHI Chapter (CHItaly 2025), October 06--10, 2025, Salerno, Italy}
\acmDOI{10.1145/3750069.3750107}
\acmISBN{979-8-4007-2102-1/25/10}

\usepackage{geometry}
\usepackage{graphicx}
\usepackage{booktabs}
\usepackage{array}
\usepackage{longtable}
\usepackage{tabularx}
\usepackage{url}
\usepackage{enumitem}
\begin{document}

\title[Am I Productive?]{Am I Productive? Exploring the Experience of Remote Workers with Task Management Tools}

\author{Russell Beale}
\authornote{RB conceptualised and supervised the work. OW designed the experiment and conducted it.  Analysis was shared. OW wrote a version of the experiment in a report; RB wrote it up into the paper}
\orcid{0000-0002-9395-1715}
\affiliation{
\department{Computer Science}\institution{University of Birmingham}
\city{Edgbaston. Birmingham}
\country{United Kingdom}}
\email{r.beale@bham.ac.uk}

\author{Oyindamola Williams}
\orcid{0009-0002-1716-9948}
\affiliation{
\department{School of Computer Science}\institution{University of Birmingham}
\city{Edgbaston, Birmingham}
\country{United Kingdom}}
\email{oyindamola@azureindigo.com}

\begin{abstract}
As the world continues to change, more and more knowledge workers are embracing remote work. Yet this comes with its challenges for their productivity, and while many Task Management applications promise to improve the productivity of remote workers, it remains unclear how effective they are. Based on existing frameworks, this study investigated the productivity needs and challenges of remote knowledge workers and how they use Task Management tools. The research was conducted through a 2-week long, mixed-methods diary study and semi-structured interview. Perceptions of productivity, task management tool use and productivity challenges were observed. The findings show that using a digital Task Management application made no significant difference to using pen and paper for improving perceived productivity of remote workers and discuss the need for better personalization of Task Management applications.
\end{abstract}

\ccsdesc[500]{Human-centered computing}
\ccsdesc[500]{Human-centered computing~Human computer interaction (HCI)}
\ccsdesc[500]{Human-centered computing~Empirical studies in HCI}
\keywords{Productivity Planning, Tool Use, Diary Study, User Psychology, Task Management, Remote Workingla}

\maketitle

\section{Introduction}

The COVID-19 pandemic precipitated a global change in how people do work. Governments and health authorities instituted lockdowns and social distancing measures that left organizations with little choice but to switch to remote work almost overnight. In April 2020, 46.6\% of workers switched to remote work~\cite{cameron2020}. For knowledge industries, the transition was relatively smooth because employees already had, or could quickly be equipped with, the necessary technology.  The prevalence of remote working in many knowledge industries spiked dramatically. In academia, for example, almost all employees were working from home. This period acted as an unplanned, large‐scale experiment in remote work. Organizations had to adapt quickly, which also exposed both the strengths (flexibility, reduced commute times) and challenges (coordination, maintaining organizational culture) of remote work.
A comprehensive review by Kniffin et al. \citeyear{Kniffin2021} highlights that the pandemic not only increased the incidence of remote work but also triggered important questions about work design, employee well-being, and long‐term organizational strategy  Remote work has now become a mainstream practice, particularly for knowledge workers who have had to adapt to working from home. Over two-thirds of employers felt at the time that they were planning to permanently shift employees to remote work, even after the Covid-19 crisis ended\cite{Kniffin2021, castrillon2020}.  

As restrictions began to ease, many organizations moved toward hybrid work models—a blend of remote and in-office work—to balance the benefits of flexibility with the need for periodic face-to-face interaction.  Studies that followed the initial pandemic response, such as those by Choudhury et al. \cite{Choudhury2021}, argue that remote work (or "work-from-anywhere" models) can maintain or even enhance productivity when employees are given geographic flexibility. These findings support the idea that many organizations may continue to embrace remote work long term, albeit in a more balanced form.The CIPD Labour Market Survey February 2025 \cite{noauthor_cipd_2025} states "Nearly half (41\%) of employers allow hybrid working with formalised policies in place".

This shift in work patterns has led many people to think about their personal productivity; what it means for them and how they can improve their task management and overall performance. There is a plethora of task management software promising to improve things for their consumers and this ever-increasing market has expanded exponentially over the past few years into a multi-billion dollar share of the software industry\cite{prodsoftmarket2020}. Previous research works in HCI have been focused on Personal Task Management, which investigates the use of task management tools and how people generally manage their tasks\cite{haraty2016}. The findings of these studies often highlight the diversity of individual preferences and differences in personal task management habits. In fact, Haraty et al. noted that a large percentage of people still prefer the pen-and-paper approach to managing their tasks.

To understand the effectiveness of Task Management (TM) software on remote workers’ productivity as they work from home, a mixed-methods diary study was carried out. The study attempts to answer the following questions:

\begin{enumerate}
    \item What are the productivity challenges and needs of remote workers?
    \item Does Task Management application software improve the perceived productivity of remote workers?
    \item How are these needs being met by the Task Management tools currently available?
\end{enumerate}

This paper is structured into 5 sections. Section 2 presents a literature review. Section 3 details the methodology used for the study. Section 4 presents the results and findings from the study. Section 5 discusses the findings and draws some conclusions. Finally, Section 6 presents the recommendations for future work.

\section{Related Work}

\subsection{Measuring Personal Productivity}

Productivity is very important to knowledge workers, although it means different things to different people. Across diverse research communities, the topic of productivity has been studied attempting to gain insights into how people work and how they can boost personal output. While there is no universally accepted definition for productivity, it is commonly defined as the relationship between output and the inputs that are required to generate that output\cite{schreyer2001}. However, this is very simplistic, especially when we consider knowledge work. The term ``knowledge worker'' is used in this paper to refer to individuals who are primarily engaged in the production, processing, or dissemination of information\cite{druryFarhoomand}. This includes, but is not restricted to, researchers, students, software developers, designers, managers, and anyone whose job involves a considerable amount of creative knowledge work.

Within the HCI research community, several studies have been conducted to understand knowledge workers’ productivity\cite{kim2019}. Within the scope of measuring productivity, two main approaches have been identified: (i) Automated measurement of product or process features to ``objectively'' assess productivity; (ii) Self-assessed productivity reporting by the individual, i.e. personal perception of productivity\cite{beller2020}. The self-assessed measure has been widely used in various studies aimed at understanding personal productivity.

In their study of knowledge workers’ productivity, Kim et al.\cite{kim2019} uncovered 6 main themes of productivity evaluation: 
\begin{enumerate}
    \item \textbf{Work product:} factors such as concrete output and progress, achievements, quality, and quantity of the outcome.
    \item \textbf{Time management:} efficiency, punctuality and use of spare time.
    \item \textbf{Attention/Distraction:} factors such as attention versus distraction, as well as emotional and physical state.
    \item \textbf{Attitude toward work:} includes mundane tasks, enjoyment and/or significance of the task, and rewards.
    \item \textbf{Impact and benefit:} long-term career benefit of the task, social or spiritual benefit, well-being and monetary reward of the task done.
    \item \textbf{Compound task:} such as multitasking, unplanned issues, and task switching.
\end{enumerate}

These themes provide some insight into a more holistic method for measuring personal productivity, especially in knowledge work which is usually far more nuanced and complicated than a simple input-output productivity measurement model. However, the study did not account for collaborative work which is common amongst many remote workers. 

The SPACE framework \cite{beller2020} proposes a more robust productivity system of measurement, including collaboration. And although it was introduced as a metric for assessing software developer productivity, it still includes variables significant enough to be applied to other forms of knowledge work. Through personal perceptions for productivity measurement, the SPACE framework measures different dimensions of productivity which are:

\begin{itemize}
    \item \textbf{Satisfaction and well-being:} including qualities such as personal satisfaction with work done, efficacy and burnout.
    \item \textbf{Performance:} measures successful process outcome (rather than output), quality and impact of the work done.
    \item \textbf{Activity:} the number of tasks and operational activities performed.
    \item \textbf{Communication and Collaboration:} how work is integrated among team members and communication effectiveness.
    \item \textbf{Efficiency and flow:} ability to get work done and complete tasks with minimal distractions or delays.
\end{itemize}

\subsection{Task Management Tools and Personal Productivity}

GTD (Getting Things Done) \cite{allen2009} is a personal productivity framework that was developed to improve productivity through focused, task-based work. With the basic principle of offloading tasks from the mind into structured task lists, the methodology is intended to reduce cognitive load and allow for greater focus. The steps that form the basis for the GTD method are:

\begin{enumerate}
    \item \textbf{Capture:} collecting every task, idea or incoming activity into an ‘inbox’.
    \item \textbf{Clarify:} defining what has been captured into clear and actionable steps.
    \item \textbf{Organize:} sorting tasks into the various projects and schedules, delegating necessary ones to any external dependencies.
    \item \textbf{Review:} regularly evaluating the task list and updating as activities progress.
    \item \textbf{Engage:} getting to work on the important tasks based on priority, context, time limit and energy levels.
\end{enumerate}

It is based on this framework that many popular task management applications have been built, with claims to boost the personal productivity of their users. However, there is little direct evidence about the effectiveness of these tools and whether they deliver on their promises. As some studies have argued, many of the commercial applications for supporting productivity and behaviour change lack a proper grounding in theoretical research \cite{stawarz2015}, both for building the applications and evaluating their effectiveness.

Applications like Trello, Asana, or Jira provide means to track work, assign responsibilities, and monitor project progression, whilst tools such as Evernote or OneNote offer powerful search, archiving, and sharing functionalities that enable knowledge capture and retrieval.  For productivity gains, however, the functionalities of these tools must integrate with workers’ routines and cognitive processes.  There is an overhead in organising and managing information that not everyone embraces \cite{jensen2018scroll}.  One pivotal study by Mueller and Oppenheimer \citeyear{mueller2014pen} demonstrated that longhand (pen-and-paper) note-taking can lead to deeper processing of information and better retention when compared with laptop note-taking. The tactile and slower nature of handwriting appears to require summarizing and processing information more actively. Pen and paper do not require power, connectivity, or updates, which makes them a reliable fallback (as discussed in works critiquing the “paperless office” myth). They provide a frictionless, distraction-free medium that is always “ready to use,” even when digital devices may interrupt work due to notifications or connectivity problems. This difference highlights a potential trade-off: while digital tools support rapid capture and integration with other systems, the analogue method may foster learning and memory consolidation. 

\subsubsection{Work Interruptions and Task Switching}
Diary studies have played an important role in understanding real-world work practices among both office and remote workers. Research by Czerwinski et al. \citeyear{czerwinski2004diary} and González and Mark \cite{gonzalez2004constant} used diary methods to record instances of task switching and interruptions. Such studies have noted that while interruptions and rapid context-switching occur in both settings, remote workers may face additional challenges such as overlapping home and work environments and asynchronous communication delay that influence productivity. These studies also help explain why some remote workers might adopt specific digital practices (or even mix digital tools with analogue methods) to maintain focus.

Remote work often requires greater self-discipline in establishing routines and boundaries. The absence of a structured physical office can lead remote workers to rely more heavily on digital diary or journaling methods to track productivity and emotional states. Such self-report methods provide researchers with granular data on work rhythms and the impact of digital interruptions. In contrast, office settings have historically been studied via observational or sensor-based methods, with diary studies becoming a useful complement in the age of remote and hybrid work.

\subsection{Why A Diary Study?}

Little is known about the efficacy of popular Task Management applications in solving the real needs of remote knowledge workers. Thus, a mixed-methods diary study and semi-structured interview was undertaken to investigate the use of these kinds of tools among users working from home.  Diary studies are qualitative (and sometimes mixed-method) approaches where participants record their activities, thoughts, and interruptions in situ. They are particularly valuable for capturing phenomena that are difficult to observe directly—such as the cognitive load imposed by switching between tasks or the subtle differences in work experience between digital and analogue mediums.  However, a disadvantage of diary studies is that they can be very tedious for the participants and even disrupt the natural flow of their daily activities\cite{ampera}. These studies have a notoriety for having declining response rates due to the seeming repetitiveness of the study’s design.

Despite these disadvantages, it was believed that this method would be useful in providing longitudinal data that would help understand the productivity behaviours, needs and challenges of remote workers and how popular tools impact their perceived productivity over time.

\section{Methodology}

To answer the research questions proposed, a mixed-methods study was carried out using a combination of diary study and semi-structured interviews to collect longitudinal data from participants.

\subsection{Preliminary Study and Screening}
Before beginning the in-depth study, a preliminary online survey was carried out. Participants were recruited from postgraduate taught students and across various social media channels. This was done to screen participants who would later take part in the longer study and to also get preliminary data that would be used in designing the longer diary study. Forty-two (n=42) participants responded to the survey; they were asked questions that included basic demographic information, use of productivity tools, productivity personality/approach and willingness to participate in the extended study. Participants who did not meet the remote work criteria or who chose not to opt in for the extended study were excluded. Out of 42 respondents, 7 were not working from home and were therefore excluded. Also, 9 respondents did not opt in for the diary study which eventually left 26 participants (74.3\%) who signed up for the 2-week long diary study.

\subsection{Experimental Design}

A between-subjects design was adopted for this study with two conditions. Participants were asked to self-describe their productivity personality, choosing between (i) thriving on planning and routine, or (ii) preferring to work in spontaneous bursts of energy.  They were then selected using stratified random sampling with 3 main strata (gender, productivity personality, and their personal productivity perception rating) into two groups, the Digital group (also called Experimental), and the Traditional group (also called Control).  The allocation balanced the different groups to make them similar to each other, for example by accounting for highly organized or particularly motivated individuals.

Participants in the Digital group were randomly assigned one of four popular digital task management tools (Todoist, Microsoft Todo, Any.do or Nirvana) which worked with the GTD framework. The Traditional group were asked to use whatever pen-and-paper form they preferred: notepads, journals, sticky notes, calendars, etc.

From our original survey that received 42 respondents, 26 remote knowledge workers (15 female, 11 male) aged between 18 and 44 volunteered to sign up for the 2-week diary study. Participation was voluntary and participants were able to withdraw at any time.  Ethical consent was obtained from the University to run the study, and participants consented to their data being collected and analysed. Eight participants dropped out of the study (one  changed their mind, one for health reasons and six were non-responsive).

Eighteen participants (10 female, 8 male) eventually participated in the full survey. They included 14 students from various institutions (77.8\%) and 4 full-time employees (22.2\%). The study ran during the April period when all would be working – four said they would be taking an Easter break; however, they were spread across the different groups and given the option to start the study later on when they would be back to work.

At this point, participants were evenly spread across the Digital and Traditional groups, with both groups having equal mean personal perceived productivity ratings (Mean = 3.22). For confidentiality, each participant was pseudonymised and assigned a unique participant ID which was used throughout the study and reporting.

\subsection{Procedure}
\subsubsection{Onboarding}
Participants were reached via their preferred communication channels (10 via Email, 7 via WhatsApp and 1 via SMS) with an initial welcome message and some general information about the study (including data protection and confidentiality notice). A pilot study was carried out for two days and notes were taken to refine the study procedure and survey questions.

Three days before the study was scheduled to start, a briefing was carried out to give participants clarity about the study. This was done via video which included a brief introduction to the survey, expectations on daily logging, specific guidelines, and instructions on task management tool sign up/set up. Participants were asked to discontinue the use of any other task management tool they were using prior to the study. Participants were encouraged to ask questions and were given clarity where needed.

\subsubsection{Data collection} The study consisted of a two-week diary study (weekdays only), followed by a post-study semi-structured interview with volunteers. The choice of a two-week study was decided as the minimum to allow familiarity with tools and give enough time to observe valid trends, behaviour change, and patterns. We recognise that a longer study would have yielded stronger results but the risk of participant dropout was felt to be too significant, and we decided  to capture deeper richer experiences over a shorter time.  For the study, participants were sent two types of messages every day: a daily reflective prompt in the morning containing guidelines on the task for the day, particularly in relation to the GTD framework, and a link to the online survey used for logging daily entries, followed by a reminder in the evening to fill in the survey.

Participants were asked to log their data entry for the day at the end of their own work day, whatever time that was - our initial survey showed widely varying work hours. Daily entries were carefully observed to ensure accuracy of data and clarity of instructions. When the response rates began to decline, participants were asked if they would prefer to change their communication channel;  4 participants were changed from Email to WhatsApp, which led to better responses. The detailed diary entry contained questions that recorded their daily productivity reflections, use of the TM tool and five main questions from the SPACE framework for measuring personal productivity. While some questions varied during the study, this productivity measure stayed consistent, as it was the dependent variable measured to test the proposed hypothesis. Using the 5-Point Likert scale where 1 represents `strongly disagree' and 5 `strongly agree'~[14], participants were asked the following:
\begin{enumerate}
    \item ``I am satisfied with the work I have done today'' (Satisfaction)
    \item ``The outcome of my work was successful'' (Performance)
    \item ``I was able to complete all of the tasks I set out for today'' (Activities)
    \item ``I collaborated successfully with my colleagues/ team members/other people'' (Collaboration and Communication)
    \item ``I was able to stay focussed on my tasks with minimal distractions'' (Efficiency and Flow)
\end{enumerate}

The responses to these questions were coded and summed to obtain an overall Productivity score.

Participants were asked questions that allowed them to rate their daily/weekly productivity levels and give open-ended answers on their reasons for this rating. They were also asked to rate their time management tool based on several parameters (e.g., reminders, scheduling, setting priorities, personalization, etc.), their perception on whether it helped improve their productivity and what they wished it did better.

The initial preliminary survey was carried out using Google Forms, the data extracted into Excel  and trends were identified for participant grouping and initial insights. MailChimp was used to send out automated emails to participants; Zoho Survey was chosen for the daily surveys.

The post-study interview with some of the participants was carried out via Zoom; interview recordings were transcribed using Otter\cite{otter}.

\section{Results}
Results are presented in two main sections. First, the quantitative results and statistical testing from the diary study are presented. Secondly, the qualitative data from the diary study and interviews are shown using thematic analysis to answer the research questions proposed.

\subsection{Quantitative Analysis}
A total of 176 entries were created by the participants for the two-week diary study. There were 4 missing entries from participants who did not fill the survey on some days.

\subsubsection{Statistical Analysis of Productivity Score}
\begin{figure}[ht]
    \centering
    \includegraphics[width=1\linewidth]{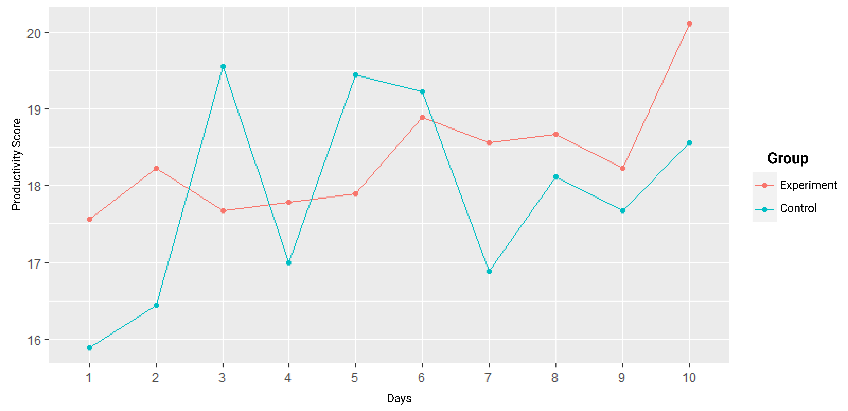}
    \caption{Self-reported productivity score for each group for 2 weeks}
    \label{fig:selfproductivityscore}
\end{figure}

The mean Productivity Score for Digital (experimental group) was 18.36 (SD = 2.57), while the mean Productivity Score for Traditional (Control group) was 17.88 (SD = 4.07). The Shapiro-Wilk test was used to check the normality of both groups (p-value = 0.0003 and 0.08416). A Two-way Mixed ANOVA was then carried out to determine the interaction between the main effects (i.e., the independent variables: productivity tool and days of work) and the dependent variable (productivity score).  The results were obtained (p = 0.887; F = 0.020) showing that there is no significant interaction, therefore we cannot reject the null hypothesis. Task management applications did not appear to improve productivity much more than traditional pen and paper.

\begin{figure}[ht]
    \centering
    \includegraphics[width=0.48\linewidth]{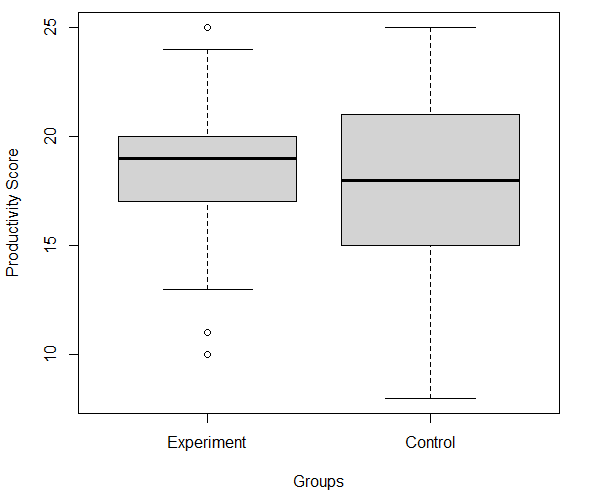}
        \includegraphics[width=0.48\linewidth]{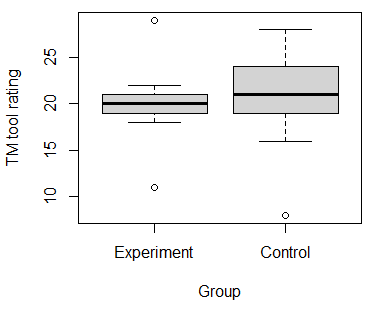}
    \caption{Left: Productivity score for each group over the study period.  Right: Task management tool rating by group}
    \label{fig:productivityscore}
\end{figure}

\subsubsection{Task Management Tool Evaluation}
The task management (TM) tool assigned to each participant group was evaluated using a technology review methodology  created by us for evaluating GTD task management software. The metrics used were versatile enough to be adapted by participants in the control group (pen and paper). They include: ease of use, reminders/notifications, customisation, interface design, reflection/tracking, and task organisation.

Responses were coded using the Likert scale and summed to obtain the TM tool rating. The possible TM tool rating ranged from 6 to 30. The null hypothesis is that the mean score for tool rating of the two groups will be the same.  The mean TM tool rating for Digital and Traditional were 20.4 and 20.1, respectively, shown in Figure \ref{fig:productivityscore} (right). The Shapiro-Wilk test was used to check the normality of both groups (p = 0.5676 and p = 0.1864). A Student T-test was performed to test the mean (-0.13406, df = 15.241, p-value = 0.8951). This shows there is no significant difference in the mean of both groups. Therefore, the we cannot reject the null hypothesis --- both the control and experimental groups have no difference in the TM tool rating.

\begin{figure}[ht]
    \centering
    \includegraphics[width=0.45\linewidth]{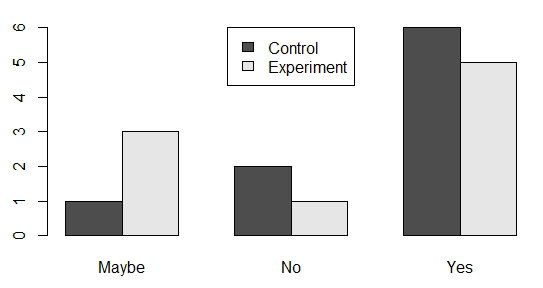}
        \includegraphics[width=0.45\linewidth]{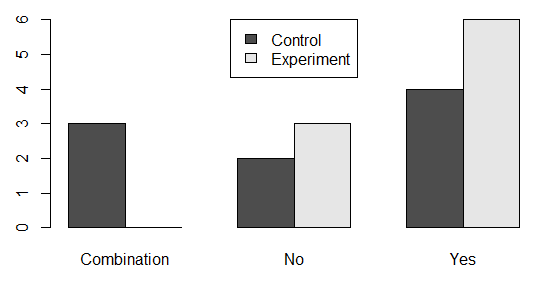}
    \caption{TM Tool effectiveness for perceived productivity by group (left), plans to continue (right)}
    \label{fig:tooleffectiveness}
\end{figure}

At the end of the study, participants were asked if they thought using the TM tool had improved their productivity by answering ``Yes'', ``No'', or ``Maybe''. The bar plot is shown in Figure \ref{fig:tooleffectiveness}. The results were analysed using the Fisher Exact test (instead of Chi-square, due to the small cell counts). The p-value = 0.6579 with a confidence level of 0.99 was obtained. This shows that there is no significant difference between the perceived TM tool effect on productivity for both the control and experimental groups.

Participants were also asked if they would wish to continue using the TM tool after the diary study. Figure \ref{fig:tooleffectiveness} (right) shows a boxplot summarising their responses. There was a variable that indicated participants wished to combine a digital TM tool with pen and paper. Statistical significance was tested using the Fisher T-test (p-value = 0.3; conf. level = 0.99).

This shows that there is no significant difference for the TM tool continuity between both the control and experimental groups.

\subsection{Qualitative Data Analysis}

After the diary study, a semi-structured interview was conducted with volunteer participants. The interviews were recorded, and a thematic analysis was performed using the inductive coding approach. In this section, the results of the qualitative data from the diary study and interviews are presented.

\subsubsection{Productivity Challenges}

Participants were asked about their biggest and current productivity challenge on Day 1 of the diary study (Table \ref{tab:challengesday1} In total, 50\% of participants answered ``Distractions'' as their biggest productivity challenge.
\begin{table*}[]
    \centering

    \begin{tabular}{m{12cm} c}
        \toprule
        Distractions: I find it hard to stay focused on my task for extended period of time because of external and/or internal distractions.
        & 50\% \\
        \midrule
        Procrastination: I sometimes put off important tasks for other less productive activities.
        & 27.8\% \\
        \midrule
        Work life balance:  I have a hard time balancing work and being present for other life priorities.
        & 5.6\% \\
        \midrule
        Prioritization: I often let urgent requests or less important tasks take over higher priority tasks.
        & 16.6\% \\
    \bottomrule
    \end{tabular}
    \caption{Perceived productivity challenges: Day 1}
    \label{tab:challengesday1}
\vspace*{-16pt}\end{table*}

\subsubsection{Thematic Analysis}

Analysing the challenges for personal productivity experienced by participants, 5 major themes were identified and summarized:


\begin{figure} 
    \includegraphics[width=0.45\textwidth]{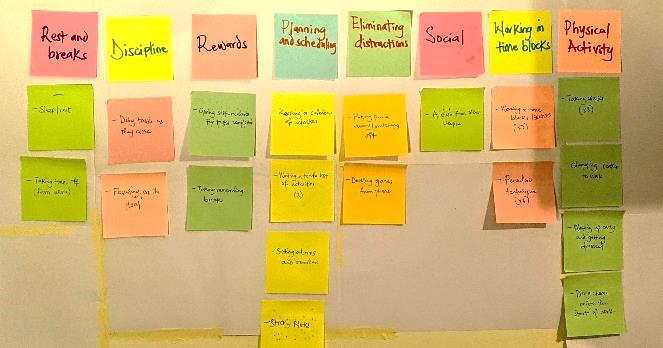}
    \caption{Sample thematic analysis}
    \label{fig:thematic analysis}
\end{figure}

\textbf{Motivation:} \\
    A major factor noted by participants was in terms of personal motivation to work. Issues such as procrastination and laziness were identified to be a problem which, in their opinion, the TM tools could not help with. In the semi-structured interview, P118 noted, \textit{"I'll be lazy again. If it's mandatory, it's fine; if they were all optional, I don't even know if I do it because I would probably sleep in the entire day."}
    
\textbf{Health and Wellbeing:} \\
    The physical and mental state of the participants was a major factor for their personal productivity. Reasons like fatigue, illness, tiredness, moods, changing energy levels, etc. were recurrent in the data. As one participant noted,\textit{ "I was super tired as I didn't sleep well, so focusing was difficult."}
    
\textbf{Distractions/Interruptions:} \\
    Participants noted distractions, unplanned events, and interruptions of various kinds as a challenge for their productivity on some days.
    
\textbf{Technical Difficulties:} \\
    Few of the participants were affected by technical issues such as power outage, which affected their ability to work.
    
\textbf{Task Estimation:} \\
    The challenge of estimating how long a task would take was a common issue raised by participants. Usually, they tended to underestimate and ended up not doing other tasks they planned. As participant 115 said, \textit{"The task required more time than I imagined."}

\subsubsection{Participant Experiences with Task Management Tools}

An analysis of responses collected from participants revealed that most of the time spent on the TM tools (digital and analogue) was spent planning tasks (i.e. identifying and writing them down) - 88\%, with 8.2\% on scheduling (setting reminders and priorities), and the remaining 3.8\% on finding information about time management, the tool, or customising it with colours, tags etc.

\begin{enumerate}
    \item \textbf{Planning and Scheduling}: A key feature of the time management that participants found useful, regardless of whether it was digital or analogue, was the ability to plan, which the tools allowed them to do. Having to think about their goals and break it down into separate tasks was an activity most participants agreed was useful for helping them try to get their tasks done.

One participant noted, `\textit{`Writing down a list of what to do just helps me to focus and actually take the time for the things I should be/want to be doing.''}

\item \textbf{Reminders and Notifications:} \\
    The major challenge which most of the control group participants experienced from using pen and paper was the inability to get reminders. One participant said this affected her productivity on a few days because she simply forgot about the task until the end of the day: \textit{``There was one task I forgot to do. Oops! I did not keep looking at the list so I forgot.''}
    
    \item \textbf{Mobility and Flexibility:} \\
    Not surprisingly, participants from the control group mentioned the immobility of pen and paper as a challenge, especially for those who had to move away from their desk at some point in their workday: \textit{``I want it to be in my face all the time lol, might look for a pocketbook in that case, so it can be really mobile'' (P111)}.
    
    \item \textbf{Satisfaction:} \\
    The satisfaction of ticking off completed tasks was noted by participants of both the control and experiment group. They explained that this motivated them to complete their other tasks. One participant noted, `\textit{`And I have a thing with lists, I really like to tick off lists, like, when I use my post it notes, I always make tick boxes beside everything. And I always get very happy to actually tick off everything, like a different coloured pen and stuff like that, just to feel that satisfaction of completing my task. So the more I started to really plan and break it down into the smaller things, I got more dependent on the app, the more useful it became for me, honestly.'' (P112)}
    
    \item \textbf{Impact on Behaviour:} \\
    Significantly, some participants across control and experiment groups noted that the TM tool did not necessarily impact their personal productivity; both self-motivated individuals who were already geared towards their goals and people needing an extra push noted that the current capabilities of commercial TM tools did not provide additional benefits. P105 noted, `\textit{`I have gotten used to doing the task and would have done it anyway.''} He often mentioned that he did not need the TM tool to complete his tasks.
    
    \item \textbf{Repeated Tasks:} \\
    The nature of the work done by remote workers sometimes requires them to have some tasks that are routine. Participants in this category noted it was repetitive to have to fill in old tasks over and over. Considering that planning tasks was the activity that took up most participants’ time, this is not surprising.\enlargethispage*{12pt}
\end{enumerate}

\subsubsection{Strategies Remote Workers Adopt for Productivity}
Participants were asked about the strategies that have helped them stay productive while working from home. Eight major themes were uncovered from the thematic analysis, shown in Table \ref{tab:strategies}.

\begin{table}[tb]
\small\begin{tabular}{@{}p{2cm} p{5cm}@{}}
\toprule
\textbf{Theme} & \textbf{Sample Responses} \\ \midrule
Discipline & ``Attending to my work as soon as they pop up…..'' --- P105 \\
 & ``Focussing on the goal'' --- P112 \\
 \midrule
Eliminating Distractions & ``I try to put away my phone sometimes'' --- P111 \\
 & ``Delete games on the phone'' --- P126 \\
 \midrule
Physical Activity & ``Getting outside in between for some fresh air and movement…'' --- P117 \\
 & ``Waking up early and getting dressed…'' --- P118 \\
 \midrule
Planning and Scheduling & ``Writing down a list of what I need to do during the week'' --- P109 \\
 & ``Setting alarms and reminders'' --- P116 \\
Rest and Breaks & ``Getting enough sleep and rest…'' --- P113 \\
 & ``Taking time off'' --- P102 \\
 \midrule
Rewards & ``…self-rewards for tasks completed.'' --- P113 \\
 & ``…take short rewarding breaks'' --- P119 \\
 \midrule
Social & ``A push from other people…'' --- P124 \\
 \midrule
Working in Time Blocks & ``I have implemented the Pomodoro technique for my work'' --- P121 \\
 & ``…breaking up the day into little sections…'' --- P117 \\
\bottomrule
\end{tabular}
\caption{Strategies Knowledge Workers Adopt for Staying Productive While Working from Home}
\label{tab:strategies}
\end{table}

\section{Discussion}

This study explored how knowledge workers use Task Management tools and their current productivity needs and challenges as they work from home. The findings show that using a digital Task Management application had no significant difference in improving the perceived productivity of remote workers compared to using a traditional pen and paper task management tool. This suggests that while participants found planning and scheduling with TM tools useful, the currently available tools are not sufficient to change their productivity behaviour or habits.

The TM tool rating, TM tool effectiveness on perceived productivity, and the TM tool continuity were also shown to have no significant difference between the experiment and control groups. One reason for this could be that most participants in the digital group were less aware of some features of the TM application they were using, focusing solely on the planning and scheduling features of the app which are more apparent. However, the simple action of planning and writing down tasks for the day was enough for most participants — whether they used a digital tool or pen and paper for this purpose. Participants who used pen and paper, however, noted that reminders would have been helpful for them, which is most likely why a high percentage of them chose to use a combination of pen and paper with a digital TM application.

\subsection{Personalization}
This study revealed that the productivity needs of everyone are different. The challenges and goals of remote workers are varied and a Task Management application that is able to account for and support these individual differences would improve the performance of users who live and work in an ever-changing context. As Haraty \citeyear{haratyPersonalized} noted, a tool’s failure to adapt to the dynamic needs of the user will often lead to users changing their tools. However, there is a danger that increased personalization could cause greater dependence on apps due to constant engagement\cite{stawarz2015} or could even distract users from actually getting their work done. Yet, with a proper implementation of context awareness, the tool should be capable of “disappearing” into the background while merging seamlessly with the user’s workflow.

\subsection{Tracking}
The ability to track progress and activity was a feature that some participants noted helped improve their productivity. Tracking involves following the progress of long-term goals or short-term tasks, as well as an automated self-monitoring feature for self-reflection, which has been shown to improve performance and influence behaviour change. This is particularly important in context because it supports the ``Review'' step of the GTD framework.

\subsection{User Experience}
Some of the productivity needs of knowledge workers highlighted in this study have already been met by some applications. For example, rescheduling unfinished tasks or setting repeat tasks are both features that participants wished their TM tool had, yet the subsequent interview revealed that some participants were unaware of these features. Although participants only had 2 weeks to use the tool it is still an issue that learning the basic functionality took that long.

\section{Limitations of the Study}
The study has many limitations. First, prompting participants to plan their day in the morning could have had an influence on their productivity perception. This was observed by some participants who said that the diary study motivated them to plan their day and helped to hold them accountable for their personal productivity. Hence, some of them might have put more effort than usual to experience higher productivity during the study period. One participant noted, \textit{"…I was more reflective of my daily tasks when filling the surveys and it made me aware of certain parts of my job that I hadn't given attention to in the past."} All participants in the different groups were exposed to the same confound.

Secondly, the participants of the study were people within the younger age bracket with advanced technological skills (which makes the lack of awareness of the capabilities of some of the digital tools even more surprising). For future studies, a wider range of individuals should be studied to get more representative data of other demographics. 

Finally, the study duration is limited: tasks and their management occur over periods of month and years and patterns in behaviour are really only seen over extended periods.  Automated quantitative metrics should be used for these longer term studies, since diary interventions are likely to exponentially decay in persistence: as we had focussed on deeper richer insights, two weeks was about the maximum we felt was feasible, and even then data completeness was dropping off.

\section{Conclusion}
This paper makes three contributions that are of interest to HCI research. First, it investigates the effect of Task Management application use on the perceived productivity of knowledge workers, and highlights their  challenges. It shows that the current tools are not significantly better than pen and paper approaches.

Through a mixed-methods study, we have shown that while the use of Task Management applications has no significant difference in the perception of productivity for knowledge workers compared to pen and paper, both tools provide great utility for helping users plan and schedule their tasks.

\bibliographystyle{ACM-Reference-Format}
\bibliography{references}
\end{document}